\documentclass[english]{revtex4}
\usepackage[T1]{fontenc}
\usepackage[latin9]{inputenc}
\usepackage{amsmath}
\usepackage{amssymb}

\makeatletter
\@ifundefined{textcolor}{}
{%
 \definecolor{BLACK}{gray}{0}
 \definecolor{WHITE}{gray}{1}
 \definecolor{RED}{rgb}{1,0,0}
 \definecolor{GREEN}{rgb}{0,1,0}
 \definecolor{BLUE}{rgb}{0,0,1}
 \definecolor{CYAN}{cmyk}{1,0,0,0}
 \definecolor{MAGENTA}{cmyk}{0,1,0,0}
 \definecolor{YELLOW}{cmyk}{0,0,1,0}
}

\makeatother

\usepackage{babel}
\begin{document}
\title{Trace Anomaly, Perelman's Functionals and the Cosmological Constant}
\author{M.J.Luo}
\address{Department of Physics, Jiangsu University, Zhenjiang 212013, People's
Republic of China}
\email{mjluo@ujs.edu.cn}

\begin{abstract}
The trace anomaly and the cosmological constant problem are two typical
breakdowns when applying the quantum principle to a general covariant
or gravitational system. A quantum theory of spacetime reference frame
is proposed and reviewed. We study the theory by functional method,
and show that the trace anomaly of the theory is closely related to
some of the Perelman's functionals. The functionals may provide us
possible links between the trace anomaly and the cosmological constant.
We find that to cancel the trace anomaly at the lab's scale up to
very high energy, a cosmological constant is required which is consistent
with observations. In the framework, an effective theory of gravity
and possible observational effect are also discussed. 
\end{abstract}
\maketitle

\section{Introduction}

Inconsistencies between two theories are the driving forces to find
a deeper one. The quantum theory and the general relativity are both
very successful in each field, but at present we are facing at least
two typical inconsistencies between them. The first one is the trace
anomaly \citep{Capper:1974ic,Deser:1976yx,Brown:1976wc,Brown:1977pq,Dowker:1976zf,Tsao:1977tj,Duff:1977ay}
when we apply the quantum principle to a general coordinates transformation
invariant system, which may indicate a breakdown of general covariance
at the quantum level. The second one is the cosmological constant
problem, (for recent reviews, see e.g. \citep{Nobbenhuis:2004wn,Bousso:2007gp,Sol2013Cosmological}),
which may indicate a serious breakdown of the equivalence principle
at the quantum level. Someone may see the problems as non-essential
side issues and can be repaired by minor patch, we consider them as
crises of current fundamental physics. Although there are some evidences
and discussions that the trace anomaly and the cosmological constant
problem are closely related \citep{Bilic:2007gr,Tomboulis:1988gw,Antoniadis:1991fa,Antoniadis:1992hz,Salehi:2000eu,Antoniadis_2007},
precisely speaking, without a consistent fundamental theory of quantum
gravity, these two problems can not be truely solved.

In previous literature \citep{Luo2014The,Luo2015Dark,Luo:2015pca,Luo:2019iby},
a framework applying quantum principle to a spacetime reference frame
is proposed, in which the general covariance and the equivalence principle
still hold and be generalized to the quantum level. So in the paper
our goal is to reconsider these two inconsistencies within the proposed
framework. In the section II, we review the fields theory of spacetime
reference frame and the induced Ricci flow. In the section III, we
derive the trace anomaly by the functional quantization approach and
find that the formal trace anomaly is related to the Shannon entropy
of the system, and a normalized trace anomaly is related to the Perelman's
partition function, which can be completely canceled at UV by a UV
limit W-entropy or related reduced volume of a shrinking Ricci soliton
solution. In the section IV, the subsequent effective gravity theory
and the cosmological constant problem is discussed. In the section
V, the observational effect of the effective gravity, for instance,
a modified distance-redshift relation at quadratic order of the redshift
is calculated which gives a deceleration parameter consistent with
observations. The conclusions of the paper is summarized in section
VI. We hope this work provide us more understandings to the mysterious
Perelman's functionals, their relations to the physical world, and
current problems of the fundamental physics.

\section{A Fields Theory of Reference Frame}

In this section, we briefly review the theory of reference frame fields
as a starting point to study a quantum theory of spacetime. The main
motivation of introducing the reference frame fields is to put a to-be-studied
system and its reference system (explicitly or implicitly used) on
an equal footing when applying the quantum principle. The treatment
of a to-be-studied system and a spacetime reference system are both
quantum, the to-be-studied quantum system is described by a state
$|\Psi\rangle$ in a Hilbert space $\mathcal{H}_{\Psi}$, and the
quantum spacetime reference system is by $|X\rangle\in\mathcal{H}_{X}$,
and the states of the whole system are given by an entangled state
\begin{equation}
|\Psi[X]\rangle=\sum\alpha_{ij}|\Psi\rangle_{i}\otimes|X\rangle_{j}
\end{equation}
in their direct product Hilbert space $\mathcal{H}_{\Psi}\otimes\mathcal{H}_{X}$.
The state $|\Psi[X]\rangle$ represents that the quantum state $\Psi$
of the to-be-studied system is reference to the quantum spacetime
coordinates $X$, replacing and generalizing the textbook formalism
$|\Psi(x)\rangle$ which can only be with respect to an inertial frame
$x$ which is classical, absolute, external and free from quantum
fluctuations. Such entangled state can only be interpreted in a relational
manner, the individual state $|\Psi\rangle$ in it has no absolute
meaning without being reference to $|X\rangle$, known as the ``non-separability
of entangled state''. In other words, the state $|\Psi\rangle$ has
physical meaning only being reference to the state $|X\rangle$ entangled
to it. When the quantum fluctuation of the reference frame $|X\rangle$
is taken into account, the state $|\Psi[X]\rangle$ can be seen as
a state with a smeared spacetime coordinate variables instead of the
textbook quantum state $|\Psi(x)\rangle$ with the classical , absolute,
external and infinitely precise coordinate variables. The state $|\Psi[X]\rangle$
could recover the textbook quantum state $|\Psi(x)\rangle$ only when
the quantum fluctuations of the reference frame state are small enough
and hence can be seen ignorable, in other words, the 2nd order (and
higher order) central moment of the spacetime coordinates $\langle\delta X^{2}\rangle$
can be ignored compared with its 1st order moment contribution $\langle X\rangle^{2}$.

To introduce a possible theory for the reference frame $X$, one way
is to consider an analogy of defining a manifolds in differential
geometry. To define a D-dimensional manifolds we need to construct
a non-linear differentiable mapping $X$ from a local coordinate patch
$x\in\mathbb{R}^{d}$ to a D-manifolds $M$. Such mapping in physics
can be realized by a kind of fields theory, the non-linear sigma model
(NLSM) \citep{gell1960axial,Friedan1980,friedan1980nonlinear,ketov2013quantum,zinn2002quantum,codello2009fixed,2015Non}
\begin{equation}
S[X]=\frac{1}{2}\lambda\int d^{d}xg_{\mu\nu}\frac{\partial X^{\mu}}{\partial x_{a}}\frac{\partial X^{\nu}}{\partial x_{a}}.\label{eq:NLSM}
\end{equation}
In a consistent formalism of a quantum fields theory, it must be formulated
in an inertial frame, on the other hand, the local coordinate patch
must also be flat, so we consider the base space $x_{a}$, $(a=0,1,...,d-1)$
of the NLSM is flat, Minkowskian or Euclidean. Considering the integral
measure $d^{d}x\equiv d^{d}x\det e$ ($\det e$ is a Jacobian) is
the same no matter the base space is in Minkowskian or Euclidean one,
since if one takes $dx_{0}^{(E)}\rightarrow idx_{0}^{(M)}$ then $\det e^{(E)}\rightarrow-i\det e^{(M)}$,
so without loss of generality, we could consider $x\in\mathbb{R}^{d}$,
and hence $d^{d}x$ is positive defined.

The differentiable mapping $X_{\mu}(x)$, $(\mu=0,1,...,D-1)$ here
is the coordinates of the target Riemannian manifolds $M$ with generally
curved metric $g_{\mu\nu}$. In the language of quantum fields theory,
$X_{\mu}(x)$ are the scalar frame fields. If we set the dimension
of the frame fields is as length $[L]$, then the prefactor $\lambda$
as a parameter of the theory has dimension of a density $[L^{-d}]$,
we will discuss its meaning and value later.

A physical interpretation of such formalism is as follows. Considering
a multi-wire proportional chamber measuring a coordinates of an event
in a lab. First we need to orient, align and order these array of
multi-wires with the reference to the wall of the lab $x_{a}$, $(a=1,2,3)$.
The electron fields in these array of multi-wires are considered as
the frame fields if we ignore their spins. The orthogonality of the
lab's wall makes the states of the electrons orthogonal to each other
and hence being independent. With the reference to the wall of the
lab, to locate the position of an event, we need to read at least
three electron signals $X_{1},X_{2},X_{3}$ in three different directions
orthogonal to each other. The information of the location can be extracted
from the wave function of these electron fields (e.g. from the phase
counting or particle number counting). Considering the electrons in
the wires are free, and the amplitude of the signals are relatively
small, so the wave function can be seen as a linear function of the
coordinates of the lab's wall, given by $X_{\mu}(x)=\sum_{a=1}^{3}e_{\mu}^{a}x_{a}$,
$(\mu=1,2,3)$ where $e_{\mu}^{a}$ in this case is proportional to
$\delta_{\mu}^{a}$ representing the amplitude of the signals in each
orthogonal direction. For the similar consideration, to know when
the event happens, we need to read an extra electron signal $X_{0}$
with the reference to the clock $x_{0}$ of the lab. So in general,
the wave functions of these 3+1 electron signals can be given by
\begin{equation}
X_{\mu}(x)=\sum_{a=0}^{3}e_{\mu}^{a}x_{a},\;(\mu=0,1,2,3),\label{eq:wavefunction of X}
\end{equation}
in which the amplitude $e_{\mu}^{a}(x)$ is nothing but the vierbein
representing a mapping from $x_{a}$ to the wave function of electron
or the spacetime frame fields $X_{\mu}(x)$.

If the event's coordinate measuring scale is beyond the scale of lab,
e.g. events happening at the distance of the scale of earth or solar
system, we could imagine to extrapolate the multi-wire chamber to
such scales without worrying about the above consideration, only replacing
the electrons by e.g. reading the photons signals. However, if the
scale is much larger than the solar system, e.g. events happening
at the distance of galaxy or even the cosmic scale, when the signal
travels along such distance scale and be read by an observer, the
broadening of the wave function of the photons or other particles
signals become unignorable. In other words, when the 2nd (or higher)
order central moment of the wave function of the signal or the spacetime
coordinate frame fields becomes important, the spacetime distance
as a quadratic form in Riemannian spacetime must be modified by the
2nd central moment (variance) fluctuation
\begin{equation}
\langle\left(\varDelta X\right)^{2}\rangle=\langle\varDelta X\rangle^{2}+\langle\delta X^{2}\rangle.
\end{equation}
Since a local distance element in Riemannian spacetime is given by
a local metric tensor at a point, so equivalently, it is convenient
to think of the location point $X$ being fixed, and we interpret
the 2nd central moment of the coordinate affect only the metric tensor
$g_{\mu\nu}$ at the location point. Then the measured value (expectation
value) of a metric tensor $g_{\mu\nu}$ is corrected by the contribution
of the 2nd central moment fluctuation of the spacetime frame fields,
\begin{equation}
\langle g_{\mu\nu}(X)\rangle=\left\langle \frac{\partial X_{\mu}}{\partial x_{a}}\frac{\partial X_{\nu}}{\partial x_{a}}\right\rangle =\frac{\partial\langle X_{\mu}\rangle}{\partial x_{a}}\frac{\partial\langle X_{\nu}\rangle}{\partial x_{a}}+\frac{1}{(\varDelta x)^{2}}\langle\delta X_{\mu}\delta X_{\nu}\rangle=g_{\mu\nu}^{(1)}(X)+\delta g_{\mu\nu}^{(2)}(X),\label{eq:g=00003Dg1+g2}
\end{equation}
where 
\begin{equation}
g_{\mu\nu}^{(1)}(X)=\frac{\partial\langle X_{\mu}\rangle}{\partial x_{a}}\frac{\partial\langle X_{\nu}\rangle}{\partial x_{a}}=\langle e_{\mu}^{a}\rangle\langle e_{\nu}^{a}\rangle\label{eq:g(1)}
\end{equation}
is the 1st order moment (average value) contribution from the classical
Riemannian spacetime. Thus even if the spacetime at the lab scale
is non-curved, the metric at large scale is in general curve due to
the fluctuation of the frame fields and hence the coordinates of the
physical spacetime. For the reason that, at the classical level, the
frame fields (\ref{eq:wavefunction of X}) given by general vierbein
satisfies the classical equation from the action of the NLSM, we call
such interpretation of NLSM a frame fields interpretation in a lab,
and the base space of NLSM is interpreted as the lab's wall and clock.
In this interpretation we could consider $d=4-\epsilon$ ($0<\epsilon\ll1$)
in (\ref{eq:NLSM}) by our common sense at least at the lab's scale
and $D\equiv4$ is the least number of the frame fields.

It is worth stressing that the setting of $d$ being not precise but
very close to 4 is for topological and quantum consistency consideration.
There are two reasons why $d$ must be very close to 4. First, certainly
at the scale of lab in our common sense, $d$ must be very close to
4. Second, if we consider the previous entangled system $\mathcal{H}_{\Psi}\otimes\mathcal{H}_{X}$,
the action of the scalar matter fields $\Psi$ (as source affecting
the spacetime frame fields and gravity) and the frame fields is
\begin{equation}
S[\Psi,X]=\int d^{d}x\left[\frac{1}{2}\frac{\partial\Psi}{\partial x_{a}}\frac{\partial\Psi}{\partial x_{a}}-\mathrm{V}(\Psi)+\frac{1}{2}\lambda g_{\mu\nu}\frac{\partial X^{\mu}}{\partial x_{a}}\frac{\partial X^{\nu}}{\partial x_{a}}\right],
\end{equation}
in which the matter field and the frame fields share the common base
space of the flat lab. The matter fields (as the to-be-studied system)
being reference to the lab frame $x$ can be transformed to be reference
to the frame fields $X$, at the semi-classical level or in 1st order
moment approximation when fluctuations of $X$ are ignored, the process
is simply a coordinates transformation $x\rightarrow X$,
\begin{align}
S[\Psi,X] & \overset{(1)}{\approx}S\left[\Psi[X]\right]=\int d^{D}X\sqrt{|\det g^{(1)}|}\left[\frac{1}{4}\left\langle g_{\mu\nu}^{(1)}\frac{\partial X^{\mu}}{\partial x_{a}}\frac{\partial X^{\nu}}{\partial x_{a}}\right\rangle \left(\frac{1}{2}g^{(1)\mu\nu}\frac{\delta\Psi}{\delta X^{\mu}}\frac{\delta\Psi}{\delta X^{\nu}}+2\lambda\right)-\mathrm{V}(\Psi)\right]\nonumber \\
 & =\int d^{D}X\sqrt{|\det g^{(1)}|}\left[\frac{1}{2}g^{(1)\mu\nu}\frac{\delta\Psi}{\delta X^{\mu}}\frac{\delta\Psi}{\delta X^{\nu}}-\mathrm{V}(\Psi)+2\lambda\right],\label{eq:semi-classical}
\end{align}
where $\overset{(1)}{\approx}$ means the 1st order moment approximation,
and $\frac{1}{4}\left\langle g_{\mu\nu}^{(1)}\frac{\partial X^{\mu}}{\partial x_{a}}\frac{\partial X^{\nu}}{\partial x_{a}}\right\rangle =\frac{1}{4}\left\langle g_{\mu\nu}^{(1)}g^{(1)\mu\nu}\right\rangle =\frac{D}{4}=1$
has been used. Thus at the semi-classical level, only consider the
1st order moment (average value), the coordinate transformation reproduces
the scalar fields action (\ref{eq:semi-classical}) in general coordinates
$X$ up to a constant $2\lambda$. $\sqrt{|\det g^{(1)}|}$ is the
Jacobian determinant of the coordinates transformation, note that
the determinant requires the coordinates transformation matrix a square
matrix, so at semi-classical level $d$ also must be very close to
$D\equiv4$. The semi-classical approximate is not valid when the
fluctuation of the frame fields is unignorable. 

The reason why $d$ not precisely 4 is as follows. For simplicity,
we consider the target spacetime of NLSM is a maximally symmetric
D-sphere, the homotopy group $\pi_{d}(S^{D=4})$ of the mapping $X(x):\mathbb{R}^{d}\rightarrow S^{D}$
is trivial for all $d<D=4$, that is to say when $d<4$ the the mapping
$X(x)$ will be free from any unphysical singularities for topological
reason, the target spacetime will always be physically well-defined.
However, $d=4$ is a little tricky, since $\pi_{4}(S^{D=4})=\mathbb{Z}$
is non-trivial, the mapping may meet intrinsic topological obstacle
and become singular. In other words, when the quantum fluctuation
of $X$ is taken into account, the RG-flow of the spacetime would
possibly develop intrinsic singularities making the theory ill-defined
and non-renormalizable (RG-flow not converge). So at the quantum level,
$d=4$ could not be precisely touched, we have to assume $d=4-\epsilon$
when the quantum principle is applying, while at the classical or
semi-classical level, it is no problem to consider $d$ being 4.

Here we discuss the frame fields beyond the semi-classical or 1st
order moment approximation. If we ignore the higher order moment fluctuations
and only consider the 2nd central moment fluctuation as the most important
next leading order contribution, we call it the Gaussian approximation
or 2nd central moment approximation. The moments higher than 2nd order
are called non-Gaussian fluctuations which are important near local
singularities of the RG-flow when the spacetime undergoes a local
phase transition, although the intrinsic global singularity may be
avoided by previous consideration. At the Gaussian approximation,
$\delta g_{\mu\nu}^{(2)}$ can be given by perturbative one-loop calculation
\citep{codello2009fixed,percacci2009asymptotic} of NLSM when it is
relatively small compared to $g_{\mu\nu}^{(1)}$
\begin{equation}
\delta g_{\mu\nu}^{(2)}=\frac{R_{\mu\nu}^{(1)}}{32\pi^{2}\lambda}\delta k^{2}
\end{equation}
where $R_{\mu\nu}^{(1)}$ is the Ricci curvature given by $g_{\mu\nu}^{(1)}$,
$k^{2}$ is the cutoff energy scale playing the role of the inverse
of the cutoff length scale of the base space $\frac{1}{(\varDelta x)^{2}}$
in (\ref{eq:g=00003Dg1+g2}). The condition of perturbative calculation
$R_{\mu\nu}^{(1)}\delta k^{2}\ll\lambda$ is valid if one notes in
later section that $\lambda$ is nothing but the critical density
of the universe, $\lambda\sim O(H_{0}^{2}/G)$ ($H_{0}\sim O(\sqrt{R})$
the current Hubble's constant and $G$ the Newton's constant). The
2nd order moment or Gaussian fluctuation depending on the cutoff scale
introduces a RG-flow to the metric tensor which is a deformation of
the spacetime driven by its Ricci curvature, known as the Ricci flow
(some reviews see e.g. \citep{chow2004ricci,chow2006hamilton,topping2006lectures})
\begin{equation}
\frac{\partial g_{\mu\nu}}{\partial t}=-2R_{\mu\nu},\;\mathrm{with}\;t=-\frac{1}{64\pi^{2}\lambda}k^{2}.\label{eq:ricci flow}
\end{equation}

There are several important features of the Ricci flow to be figured
out. First, the Ricci flow equation is non-linear, and along the flow
parameter $t$, the Ricci flow is an averaging or coarse graining
process to the non-linear gravitational system which is highly non-trivial
\citep{carfora1995renormalization,piotrkowska1995averaging,carfora2008ricci,Zalaletdinov:2008ts,Paranjape:2009zu}.
If the flow is free from local singularities then there exists long
flow-time solution in $t\in(-\infty,0)$, (it is often called an ancient
solution in literature). This range of $t$ corresponds to $k\in(\infty,0)$,
i.e. from short distance (high energy) scale UV flows forwardly to
long distance (low energy) scale IR. The metric at certain scale $t$
is given by having averaged out the shorter distance details which
produces an effective correction to the effective metric at the scale.
So as the flow flows to IR, it losses its information in shorter distance,
so the flow is non-reversible. Second, the 2nd order central moment
fluctuation modifies the local distance of a spacetime which is not
important for its topology, so along the Ricci flow parameter $t$,
it preserves the topology of the spacetime but its local metric, shape
and volume change. There is a special solution of the Ricci flow called
Ricci soliton which only changes its local volume while keeps its
local shape. The Ricci soliton as a generalization of the notion of
fixed point of the Ricci flow is an important concept in understanding
the large scale cosmological observations in our following sections.

The Ricci flow was introduced in 1980s by Friedan \citep{Friedan1980,friedan1980nonlinear}
in physics and independently by Hamilton in mathematics \citep{Hamilton1982Three,hamilton1986four}.
The main motivation of introducing the Ricci flow from the mathematical
point of view is to classify manifolds, a specific goal is to proof
the Poincare conjecture. Hamilton used it as a tool to gradually deform
a manifolds into a more and more ``simple and good'' manifolds or
pieces whose topology can be easily recognized. The program was fully
realized owing to Perelman's breakthrough around 2003 \citep{perelman2002entropy,perelman2003ricci,perelman307245finite}
by introducing several monotonic functionals to successfully deal
with the local singularities which may be developed under the Ricci
flow. From physics point of view, the classification of a manifolds
is equivalent to constructing the complete Hilbert space or the phase
diagram of the spacetime, which means a quantization of the spacetime.
For example, the approach of the Ricci flow with surgery \citep{perelman2003ricci}
proved the Thurston's conjecture that the 3-manifolds has only 8 topologically
inequivalent types of fundamental building blocks, which in the language
of physics is equivalent to claim that the dimension of the Hilbert
space of the 3-space is finite. We could consider the Ricci flow at
least at the Gaussian approximation level represents a possible quantum
theory of spacetime. Via the assumption that the Equivalent Principle
of Einstein can be consistently generalized to the quantum level,
the Ricci flow can also be interpreted as a possible quantum theory
of gravity. The correctness of these claims depends on the mathematical
consistency of the theory and the validity in applying it to explain
and predict observations.

To handle the difficult Ricci flow equation, there is a trick in mathematical
literature that introduces a density function $u(X,t)$ to cancel
the flow of the volume element $d^{D}X$ or $\sqrt{|\det g|}$ of
the target spacetime, i.e.
\begin{equation}
\frac{\partial}{\partial t}\left(u\sqrt{|\det g|}\right)=0.\label{eq:volume constraint-1}
\end{equation}
The density $u$ modifies the Ricci flow by a family of diffeomorphism
equivalent to the standard Ricci flow, gives rise to the Ricci-DeTurck
flow \citep{deturck1983deforming}
\begin{equation}
\frac{\partial g_{\mu\nu}}{\partial t}=-2R_{\mu\nu}+2\nabla_{\mu}\nabla_{\nu}\log u.\label{eq:ricci-deturck}
\end{equation}
By using (\ref{eq:volume constraint-1}) and (\ref{eq:ricci-deturck})
we have a flow equation of density $u$
\begin{equation}
\frac{\partial u}{\partial t}=-\left(\Delta-R\right)u,\label{eq:u-flow t}
\end{equation}
where $\Delta$ is the Laplace-Beltrami operator of the target D-spacetime.
The equation (\ref{eq:u-flow t}) is a backwards heat-like equation.
Naively speaking, the solution of the backwards heat flow will not
be exist. But one of the basic fact is that if one let the Ricci flow
flow to certain IR scale $t_{*}$, and at $t_{*}$ one may then choose
an appropriate $u(t_{*})=u_{0}$ arbitrarily (different choices of
$u$ is equivalent to the choice of a diffeomorphism gauge) and flow
it backwards in $\tau=t_{*}-t$ to obtain a solution $u(\tau)$ of
the backwards equation. If the flow is free from singularities, e.g.
for a homogeneous and isotropic flow, we could simply choose $t_{*}=0$,
so we defined
\begin{equation}
\tau=0-t=\frac{1}{64\pi^{2}\lambda}k^{2}\in(0,\infty).\label{eq:tau}
\end{equation}
In this case, the density $u(X,\tau)$ satisfies the heat-like equation
\begin{equation}
\frac{\partial u}{\partial\tau}=\left(\Delta-R\right)u,\label{eq:u-flow}
\end{equation}
which does admit a solution along $\tau$. At this point, the mathematical
problem of the Ricci flow of a Riemannian manifolds is transformed
to coupled equations
\begin{equation}
\begin{cases}
\begin{array}{c}
\frac{\partial g_{\mu\nu}}{\partial t}=-2R_{\mu\nu}+2\nabla_{\mu}\nabla_{\nu}\log u\\
\frac{\partial u}{\partial\tau}=\left(\Delta-R\right)u\\
\frac{d\tau}{dt}=-1
\end{array}\end{cases}
\end{equation}
and the pure Riemannian manifolds $(M,g)$ is transformed to a Riemannian
density manifolds $(M,g,u)$ \citep{Morgan2009Manifolds,2016arXiv160208000W,Corwin2017Differential}
with the constraint (\ref{eq:volume constraint-1}) .

The density $u$ plays a particular important role in studying the
Ricci flow and the possible theory of quantum spacetime because, first
it introduces a fixed measure along the flow which is related to the
interpretation of the base space of NLSM as the observer's fiducial
reference (the lab's wall and clock); second, it produces the Ricci-DeTurck
flow equivalent to the Ricci flow up to a family of diffeomorphism,
which turns out to be a gradient flow of some Perelman's functionals;
third, the density $u$ has deep statistic and thermodynamics interpretation
and implications for a quantum spacetime, making a deep understanding
of a microscopic theory of quantum spacetime possible; and fourth,
in the spirit of comparison geometry, the density $u$ encodes the
most crucial geometric feature of a flowing Riemannian spacetime,
i.e. the local volume ratio between the fiducial volume of an observer's
lab and the one at a distant spacetime, making the formulation of
the trace anomaly, Perelman's functionals, and some physical quantities
of gravity in the following sections much convenient.

\section{Quantization of the Frame Fields and the Trace Anomaly}

As mentioned in the previous section that the Ricci flow as a RG-flow
of the NLSM does not preserve the volume of a spacetime, even if the
volume of the lab is fixed (\ref{eq:volume constraint-2}). Especially
at a flow limit: the Ricci soliton, the Ricci flow only changes its
volume but shape. As a consequence, a trace anomaly due to this part
of degrees of freedom of spacetime may emerge when we apply the quantum
principle to the system. In the section, we consider the quantization
of the pure frame fields by using the functional method and deduce
the trace anomaly. First we define the partition function by the functional
integral with the NLSM action (\ref{eq:NLSM})
\begin{equation}
Z=\int[\mathcal{D}X]\exp\left(-S[X]\right)=\int[\mathcal{D}X]\exp\left(-\frac{1}{2}\lambda\int d^{d}xg_{\mu\nu}\partial_{a}X^{\mu}\partial_{a}X^{\nu}\right),
\end{equation}
which is independent to whether the base space is Euclidean or Minkowskian.

Note that the general coordinates transformation
\begin{equation}
\hat{X}_{\mu}=e_{\mu}^{\nu}X_{\nu}
\end{equation}
does not change the action $S[X]$, but the measure of functional
integral $\mathcal{D}\hat{X}$ changes,
\begin{align}
\mathcal{D}\hat{X} & =\prod_{x}\prod_{\mu=0}^{D}d\hat{X}_{\mu}(x)=\prod_{x}\epsilon_{\mu\nu\rho\sigma}e_{\mu}^{0}e_{\nu}^{1}e_{\rho}^{2}e_{\sigma}^{3}dX_{0}(x)dX_{1}(x)dX_{2}(x)dX_{3}(x)\nonumber \\
 & =\prod_{x}\left|\det e(x)\right|\prod_{x}\prod_{a=0}^{D}dX_{a}(x)\nonumber \\
 & =\left(\prod_{x}\left|\det e\right|\right)\mathcal{D}X
\end{align}
where
\begin{equation}
\epsilon_{\mu\nu\rho\sigma}e_{\mu}^{0}e_{\nu}^{1}e_{\rho}^{2}e_{\sigma}^{3}=\left|\det e\right|=\sqrt{\left|\det g\right|}
\end{equation}
is the Jacobian of the vierbein at point $x$. The Jacobian is nothing
but the local relative volume element w.r.t. the fiducial frame $X$,
which also represents local longitudinal volume change degrees of
freedom of spacetime. If we consider the volume element $d^{4}x$
of the lab is fixed,
\begin{equation}
\int_{M}\left|\det e\right|^{-1}d^{D}X=\lambda^{-1}=\int d^{4}x,\label{eq:volume constraint-2}
\end{equation}
so the Jacobian is just a local density $u$ of the spacetime, 
\begin{equation}
u(x)\equiv\frac{d^{4}x}{d^{D}X(x)}=\left|\det e\right|^{-1}.
\end{equation}
The solution $u$ of (\ref{eq:u-flow}) can be parameterized as
\begin{equation}
u=\frac{1}{(4\pi\tau)^{D/2}}e^{-f}.\label{eq:u}
\end{equation}
Therefore the partition function now is given by
\begin{align}
Z(\hat{M}) & =\int[\mathcal{D}\hat{X}]\exp\left(-S[\hat{X}]\right)=\int\left(\prod_{x}\left|\det e\right|\right)[\mathcal{D}X]\exp\left(-S[X]\right)\nonumber \\
 & =\int\left(\prod_{x}e^{f+\frac{D}{2}\log(4\pi\tau)}\right)[\mathcal{D}X]\exp\left(-S[X]\right)\nonumber \\
 & =\int[\mathcal{D}X]\exp\left\{ -S[X]+\lambda\int d^{4}x\left[f+\frac{D}{2}\log(4\pi\tau)\right]\right\} \nonumber \\
 & =\int[\mathcal{D}X]\exp\left\{ -S[X]+\lambda\int_{M}d^{D}Xu\left[f+\frac{D}{2}\log(4\pi\tau)\right]\right\} 
\end{align}
Now we see, at the quantum level, there is a change of the partition
function and hence if one deduces the stress tensor by $\langle T_{\mu\nu}\rangle=\frac{\delta\log Z}{\delta g^{\mu\nu}}$,
its trace $\langle g^{\mu\nu}\rangle\langle T_{\mu\nu}\rangle$ is
different from $\langle T_{\mu}^{\mu}\rangle=\langle g^{\mu\nu}T_{\mu\nu}\rangle$,
i.e.
\begin{equation}
\langle\varDelta T_{\mu}^{\mu}\rangle=\langle g^{\mu\nu}\rangle\langle T_{\mu\nu}\rangle-\langle g^{\mu\nu}T_{\mu\nu}\rangle=\lambda\left[f+\frac{D}{2}\log(4\pi\tau)\right]\neq0
\end{equation}
known as the trace anomaly. It comes from the non-invariance of the
partition function
\begin{equation}
Z(M)\rightarrow Z(\hat{M})=e^{\lambda\int_{M}d^{D}Xu\left[f+\frac{D}{2}\log(4\pi\tau)\right]}Z(M),
\end{equation}
which means a breakdown of the general covariance at the quantum level.
The trace anomaly is purely because the longitudinal degrees of freedom
of spacetime do not preserve the functional integral measure which
change the spacetime volume. 

Note that the trace anomaly of the action is nothing but the Shannon
entropy of the spacetime in terms of the density $u$ playing the
role of a (to be normalized) probability density,
\begin{equation}
\int_{M}d^{D}Xu\left[f+\frac{D}{2}\log(4\pi\tau)\right]=-\int_{M}d^{D}Xu\log u=N.
\end{equation}
In order to normalize the density $u$ or equivalently apply the first
equal sign of the constraint (\ref{eq:volume constraint-2}) to the
$f$ function, consider a Gaussian type density
\begin{equation}
u_{G}=\frac{1}{(4\pi\tau)^{D/2}}e^{-\frac{|X|^{2}}{4\tau}},
\end{equation}
which is also a solution of (\ref{eq:u-flow}) and satisfies the constraint
(\ref{eq:volume constraint-2}). Its Shannon entropy is
\begin{equation}
N_{G}=-\int_{M}d^{D}Xu_{G}\log u_{G}=\int d^{4}x\frac{D}{2}\left[1+\log(4\pi\tau)\right]=\int_{M}d^{4}Xu\frac{D}{2}\left[1+\log(4\pi\tau)\right].
\end{equation}
Thus a normalized $f$ function gives a normalized Shannon entropy
which is just the -log of the Perelman's partition function \citep{perelman2002entropy}
\begin{equation}
\tilde{N}(M)=N-N_{G}=\int_{M}d^{D}Xu\left(f-\frac{D}{2}\right)\equiv-\log Z_{P}.\label{eq:Ntilde}
\end{equation}
The normalized Shannon entropy $\tilde{N}$ is defined to be zero
at an equilibrium state when the density $u$ is an equilibrium Gaussian
type at the flow limit
\begin{equation}
\lim_{\tau\rightarrow0}f(X,\tau)=\frac{|X|^{2}}{4\tau}.\label{eq:f at tau=00003D0}
\end{equation}
And more precisely, the normalized Shannon entropy is zero at a final
IR fixed point of the Ricci flow
\begin{equation}
\lim_{\tau\rightarrow0}\tilde{N}(M)=0.\label{eq:Ntilde at tau=00003D0}
\end{equation}
The monotonicity of $\tilde{N}(M,\tau)$ implies $\tilde{N}\le0$
and the equality can only be taken at $\tau\rightarrow0$, which is
simply because the equilibrium entropy $N_{G}$ gives the maximal
entropy. Thus at this point, the normalized Shannon entropy as the
trace anomaly is gone at a final IR fixed point of the flow, but in
general scale the trace anomaly is intrinsic. 

The partition function now takes the form
\begin{equation}
Z(\hat{M})=e^{\lambda\tilde{N}(M)}Z(M)=\frac{Z(M)}{Z_{P}^{\lambda}(M)}.
\end{equation}

As is pointed out in \citep{Deser:1993yx} that there are generally
two types of geometric origin of the trace anomalies: Type-A comes
from the non-invariance of the functional integral measure summing
over non-dynamical topological degrees of freedom in all possible
different topological equivalent classes which contributes a scale
invariant topological term to its action; and Type-B comes from also
the non-invariance of the functional integral measure summing over
the dynamical longitudinal degrees of freedom in a single topological
equivalent class which is scale dependent. The Ricci flow preserves
the topology of the initial manifolds until the flow encounters intrinsic
singularity, in principle all local non-intrinsic (finite flow time)
singularities of the flow could be carefully removed by performing
proper local surgeries \citep{perelman2003ricci,perelman307245finite}
(which is generally believed valid not only for 3-manifolds). The
choice of $d=4-\epsilon$ also guarantees the absence of an intrinsic
topological obstacle in the mapping $x\rightarrow X$. Thus the density
$u$ can be seen always within a single topological equivalent class
along its backwards flow, and never cross to another gauge inequivalent
regime. In other words, the gauge is fixed within a particular topological
equivalent class depending on the initial topology of the spacetime,
so that along the flow, the volume of the spacetime and hence the
functional integral measure changes but its global topology of the
spacetime is always preserved, there is no extra contributions to
the functional integral measure from other topological inequivalent
classes. So in this sense the trace anomaly obtained here is of the
Type-B which requires a scale $\tau$, and a counter term canceling
the anomaly obviously also depends on the initial topology of the
spacetime depicted by a kind of topological invariant.

As we have imposed the second equal sign of the constraint (\ref{eq:volume constraint-2})
and hence the volume of the lab must be fixed, if we need a consistency
between the general coordinates transformation invariance and the
quantum principle, the trace anomaly must be canceled at the starting
reference: the lab's scale which is up to a very high energy UV scale.
The cancellation of the trace anomaly at UV is equivalent to set a
proper initial condition for the volume flow starting from UV.

In order to see how the anomaly can be canceled at UV, first we note
that the trace anomaly term is monotonic along $\tau$ and has many
analogous behavior with the thermodynamic functions. For the reason,
Perelman introduced his W-functional \citep{perelman2002entropy}
\begin{equation}
\mathcal{W}(M,u,\tau)=\tau\tilde{\mathcal{F}}+\tilde{N}=\frac{d}{d\tau}\left(\tau\tilde{N}\right),\label{eq:W}
\end{equation}
where 
\begin{equation}
\lambda\tilde{\mathcal{F}}=\lambda\mathcal{F}-\frac{D}{2\tau}=\lambda\int d^{D}Xu\left(R+|\nabla f|^{2}\right)-\frac{D}{2\tau}\label{eq:Ftilde}
\end{equation}
is the equilibrium normalized F-functional of Perelman \citep{perelman2002entropy}.
When we use the analogies \citep{perelman2002entropy} e.g. the temperature
$T\sim\tau$ , the internal energy $-\tau^{2}\lambda\tilde{\mathcal{F}}\sim U$,
the thermodynamic entropy $S\sim-\lambda\mathcal{W}$, and the free
energy $F\sim\lambda\tau\tilde{N}$, the equation (\ref{eq:W}) is
in analogy to the thermodynamic equation $U-TS=F$. So in this sense
the W-functional of Perelman is usually called the W-entropy. The
monotonicity of $\tilde{N}$ and the W-functional implies that they
converge to a constant $\nu(M_{\infty})$ at UV
\begin{equation}
\lim_{\tau\rightarrow\infty}\lambda\tilde{N}(M,u,\tau)=\lim_{\tau\rightarrow\infty}\lambda\mathcal{W}(M,u,\tau)=\inf_{\tau}\lambda\mathcal{W}(M,u,\tau)=\nu(M_{\infty})<0,
\end{equation}
under the normalized condition (\ref{eq:volume constraint-2}). So
the trace anomaly can be completely canceled at UV by introducing
the constant $\nu(M_{\infty})$,
\begin{equation}
Z(\hat{M})=e^{\lambda\tilde{N}(M)-\nu(M_{\infty})}Z(M)=e^{\lambda\int_{M}d^{D}Xu\left[f-\frac{D}{2}-\nu\right]}Z(M)\label{eq:partition}
\end{equation}

The counter term $e^{\nu}$ is usually called the Gaussian density
\citep{cao2004gaussian,cao2009recent} of an ancient ($\tau$ could
trace back to UV $\tau\rightarrow\infty$ without encountering any
singularities) manifolds $M$ in the literature of Ricci flow. For
a maximally symmetric spacetime that space and time are on an equal
footing, the transverse modes which change the shape of the spacetime
damp, leaving only the longitudinal modes which only change the volume
or size of the spacetime so that it can go back to UV $\tau\rightarrow\infty$.
The Gaussian density \citep{xu2017equation} of such spacetime can
be given by the calculation of the UV limit of the reduced volume
$\tilde{V}(M_{\infty})$ introduced also by Perelman
\begin{equation}
e^{\nu}=\tilde{V}(M_{\infty})=\lim_{\tau\rightarrow\infty}\int_{M}d^{D}Xv(X,\tau)\label{eq:RV}
\end{equation}
where 
\begin{equation}
v(X,\tau)=\frac{1}{(4\pi\tau)^{D/2}}e^{-l(X,g,\tau)}
\end{equation}
is a subsolution \citep{perelman2002entropy} of the conjugate heat
equation (\ref{eq:u-flow}), i.e.
\begin{equation}
\left(\frac{\partial}{\partial\tau}-\Delta+R\right)v\le0.\label{eq:v_flow}
\end{equation}
 $l(X,g,\tau)$ is the reduced length \citep{perelman2002entropy}
measuring the minimum distance of a path between the base space point
$\gamma(0)=0$ and an end point $\gamma(\tau)=X$
\begin{equation}
l(X,g,\tau)=\inf_{\gamma}\frac{1}{2\sqrt{\tau}}\int_{0}^{\tau}\sqrt{\tau^{\prime}}\left[R\left(\gamma(\tau^{\prime})\right)+\left|\frac{d\gamma}{d\tau^{\prime}}\right|_{g(\tau^{\prime})}^{2}\right]d\tau^{\prime}\label{eq:reduced length}
\end{equation}
where $R\left(\gamma(\tau^{\prime})\right)$ is the scalar curvature
at the point $\gamma(\tau^{\prime})$ and ``$\inf$'' (infimum)
is taken over all path $\gamma$ with the fixed base and end points.
The inequality in the (\ref{eq:v_flow}) holds as equality only when
the manifolds $M$ is a gradient shrinking soliton.

As an example, now we calculate the reduced volume of a maximally
symmetric D-ball with curvature radius $a(\tau\rightarrow\infty)$.
In fact, the action is independent to whether the base space is Euclidean
or Minkowskian, compact or non-compact, the metric now is given by
\begin{equation}
g_{\mu\nu}^{(E)}(\tau)=a^{2}(\tau)\delta_{\mu\nu}\quad\mathrm{or}\quad g_{\mu\nu}^{(M)}(\tau)=a^{2}(\tau)\eta_{\mu\nu}.
\end{equation}
By using the Ricci flow equation (\ref{eq:ricci flow}) with $t=-\tau$
and the Ricci curvature $R_{\mu\nu}=\frac{D-1}{a^{2}}g_{\mu\nu}$,
we have the flow of the radius is
\begin{equation}
a^{2}(\tau)=2(D-1)\tau,
\end{equation}
as $\tau\rightarrow0$ the radius shrinks to a singular point. In
this case, note that the Ricci flow equation is nothing but a shrinking
soliton equation,
\begin{equation}
R_{\mu\nu}=\frac{D-1}{a^{2}}g_{\mu\nu}=\frac{1}{2\tau}g_{\mu\nu}.
\end{equation}
In other words, the Ricci flow of the maximally symmetric spacetime
is a shrinking soliton which only shrinks the volume of the D-ball
as $\tau\rightarrow0$ without changing its shape. More over, when
tracing back to the UV limit $\tau\rightarrow\infty$, we have $a(\tau)\rightarrow\infty$,
so the scalar curvature in (\ref{eq:v_flow}) tends to zero, and hence
the reduced length (\ref{eq:reduced length}) tends to a Gaussian
distance $|X|^{2}/4\tau$. As a consequence, the reduced volume (\ref{eq:RV})
can be obtained by using a standard heat kernel $\frac{1}{(4\pi\tau)^{D/2}}e^{-\frac{|X|^{2}}{4\tau}}$
to approximate the $v(X,\tau)$ function 
\begin{align}
e^{\nu}=\tilde{V}(B_{\infty}) & \approx\lim_{\tau\rightarrow\infty}\frac{1}{(4\pi\tau)^{D/2}}\int_{B_{\tau}}d^{D}Xe^{-\frac{X^{2}}{4\tau}}\nonumber \\
 & =\lim_{\tau\rightarrow\infty}\frac{1}{(4\pi\tau)^{D/2}}\int_{0}^{a(\tau)=\sqrt{2(D-1)\tau}}e^{-\frac{r^{2}}{4\tau}}\frac{D\pi^{D/2}}{\Gamma\left(\frac{D}{2}+1\right)}r^{D-1}dr\nonumber \\
 & \overset{D=4}{=}0.442\label{eq:e^nu}
\end{align}
Since when the spacetime is flat, the volume of $\mathbb{R}^{D}$
is given by
\begin{equation}
\lim_{\tau\rightarrow\infty}\int_{\mathbb{R}^{D}}d^{D}Xe^{-\frac{X^{2}}{4\tau}}=\lim_{\tau\rightarrow\infty}\int_{0}^{\infty}e^{-\frac{r^{2}}{4\tau}}\frac{D\pi^{D/2}}{\Gamma\left(\frac{D}{2}+1\right)}r^{D-1}dr=\lim_{\tau\rightarrow\infty}(4\pi\tau)^{D/2},
\end{equation}
so the reduced volume in fact measures a relative volume or volume
ratio between a ball-volume of curvature radius $a$ and volume of
a fiducial flat $\mathbb{R}^{D}$ spacetime (e.g. the lab frame of
an observer),
\begin{equation}
0<e^{\nu}=\tilde{V}(M_{\infty})=\frac{\mathrm{Vol}(M_{\infty}^{D})}{\mathrm{Vol}(\mathbb{R}^{D})}<1.
\end{equation}

Because the spacetime is shrinking w.r.t. the fiducial volume when
the curvature is positive, the volume ratio is clearly less than 1.
In other words, since the counter term is just the initial value of
the $u$ density, $e^{\nu}=\tilde{V}(M_{\infty})=u_{\tau=0}^{-1}$,
an observer's spacetime of unit volume from UV flows and converges
to a finite $u_{0}$ and hence a finite relative volume $0<e^{\nu}<1$
instead of shrinking to the singular point at IR $\tau\rightarrow0$.
The trace anomaly $e^{\lambda\tilde{N}(M)}$ is completely canceled
by $e^{\nu}$ at UV, making the partition function $Z(\hat{M}_{\infty})=Z(M_{\infty})$
in the observer's lab frame invariant under the general coordinates
transformation at the quantum level.

\section{Effective Theory of Gravity}

In the previous section, we can see the partition function (\ref{eq:partition})
can be factorized into the fiducial part $Z(M)$ and the pure longitudinal
part $e^{-\nu}Z_{P}^{-\lambda}$, in which $Z(M)$ comes from the
transverse degrees of freedom of the fiducial spacetime changing the
shape of the spacetime and matters as the source of gravity, $Z_{P}^{\lambda}$
is given by the Perelman's partition function, and $e^{\nu}$ is calculated
by a shrinking Ricci soliton configuration because of the pure volume-changing
and longitudinal nature. For simplicity, we could consider the lab
frame as the pure fiducial spacetime without matter coupling to it
(the general matter coupling situation can be found in \citep{Luo:2015pca,Luo:2019iby}).
Since the lab frame is interpreted as the base space and hence can
be treated classically, so we have the fiducial partition function

\begin{equation}
Z(M)=\exp\left(-\frac{1}{2}\lambda\int d^{4}xg_{\mu\nu}\partial_{a}X^{\mu}\partial_{a}X^{\nu}\right)=\exp\left(-\frac{1}{2}\lambda\int d^{4}xg_{\mu\nu}g^{\mu\nu}\right)=\exp\left(-\int d^{4}x\frac{D}{2}\lambda\right)=e^{-S_{cl}}.
\end{equation}
Therefore, in the situation, the whole partition function is 
\begin{equation}
Z(\hat{M})=e^{\lambda\tilde{N}(M)-\nu}Z(M)=\exp\left[-S_{cl}+\lambda\tilde{N}(M)-\nu\right]=\exp\left[\lambda\int_{M}d^{4}Xu\left(f-D\right)-\nu\right].\label{eq:exact action}
\end{equation}
 Because $\lim_{\tau\rightarrow0}\tilde{N}(M)=0$, so at small $\tau$,
$\tilde{N}(M)$ can be expanded by powers of $\tau$, we have
\begin{equation}
\tilde{N}(M)=\frac{\partial\tilde{N}}{\partial\tau}\tau+O(\tau^{2})=\int d^{D}Xu\left[\left(R_{\tau=0}+|\nabla f_{\tau=0}|^{2}-\frac{D}{2\tau}\right)\tau\right]+O(\tau^{2})=\int d^{D}XuR_{0}\tau+O(\tau^{2}),
\end{equation}
in which $\frac{\partial\tilde{N}}{\partial\tau}=\tilde{\mathcal{F}}$
in (\ref{eq:Ftilde}) , eq.(\ref{eq:f at tau=00003D0}) and hence
$\lambda\int d^{D}Xu\tau|\nabla f_{\tau=0}|^{2}=\frac{D}{2}$ has
been used. Then the partition function can be approximately given
by
\begin{align}
Z(\hat{M}) & =e^{-S_{cl}-S_{k}}\approx\exp\left[-S_{cl}+\int_{M}d^{D}Xu\left(\lambda R_{0}\tau-\lambda\nu\right)\right],\;(\mathrm{small}\:\tau)
\end{align}
by taking (\ref{eq:tau}), the effective action at energy cutoff scale
$k$ is given by
\begin{equation}
S_{cl}+S_{k}\approx\int_{M}d^{4}X\sqrt{|\det g|}\left(2\lambda-\frac{R_{0}}{64\pi^{2}}k^{2}+\lambda\nu\right),\;(\mathrm{small}\:k)\label{eq:effective action}
\end{equation}
which is only a low energy approximation of the exact action (\ref{eq:exact action}).

We consider (\ref{eq:exact action}) as the action of a pure gravity,
so it must recover the Einstein-Hilbert (EH) action when the scale
$k$ ranges from the lab scale to the solar system scale (where $\tau$
is away from 0) where the Einstein's theory of gravity is well tested.
However, at the cosmic scale ($\tau\rightarrow0$) we know that the
EH action deviates from observation, where the cosmological constant
becomes important. 

For one thing, note that at the cosmic scale $\tau\rightarrow0$,
$\tilde{N}(M)\rightarrow0$, then the action leaving the fiducial
partition function $Z(M)=e^{-S_{cl}}$ with Lagrangian $\mathcal{L}_{cl}=\frac{D}{2}\lambda$
plus $\lambda\nu$, i.e. $2\lambda+\lambda\nu$ term should play the
role of the standard EH action with a constant scalar curvature $R_{0}$
plus the cosmological constant term, i.e.
\begin{equation}
\mathcal{L}_{cl}+\lambda\nu=2\lambda+\lambda\nu=\frac{R_{0}-2\Lambda}{16\pi G}.\label{eq:EH at k=00003D0}
\end{equation}

For another thing, because at UV, $\tau\rightarrow\infty$, $\lambda\tilde{N}(M)\rightarrow\nu$,
the action leaving only the fiducial Lagrangian $\mathcal{L}_{cl}=2\lambda$
which should be interpreted as a constant EH action without the cosmological
constant, i.e.
\begin{equation}
\mathcal{L}_{cl}=2\lambda=\frac{R_{0}}{16\pi G}.
\end{equation}
So we have the cosmological term
\begin{equation}
\lambda\nu(M_{\infty})=\frac{-2\Lambda}{16\pi G}=-\rho_{\Lambda}.
\end{equation}
A major difference between the effective action (\ref{eq:effective action})
and the standard EH action is that this action has gradient flow but
the standard EH action does not. The first two terms in (\ref{eq:effective action})
can be considered as an effective EH Lagrangian
\begin{equation}
2\lambda-\frac{R_{0}}{64\pi^{2}}k^{2}=\frac{R_{k}}{16\pi G},\;(\mathrm{small}\:k)
\end{equation}
which is nothing but the flow of the scalar curvature \citep{chow2007ricci}
\begin{equation}
R_{k}=\frac{R_{0}}{1+\frac{1}{4\pi}Gk^{2}},\quad\mathrm{or}\quad R_{\tau}=\frac{R_{0}}{1+\frac{2}{D}R_{0}\tau}.
\end{equation}

We see that at the cosmic scale $k\rightarrow0$ the effective scalar
curvature is bounded by $R_{0}$ which can be obtained by measuring
the ``Hubble's constant'' $H_{0}$ at the cosmic scale,
\begin{equation}
R_{0}=D(D-1)H_{0}^{2}=12H_{0}^{2}.
\end{equation}
So it is very surprise to note that $\lambda$ is nothing but the
critical density of the universe
\begin{equation}
\lambda=\frac{3H_{0}^{2}}{8\pi G}=\rho_{c},
\end{equation}
so the cosmological constant term is of the order of the scale of
the critical density with a ``dark energy'' fraction
\begin{equation}
\Omega_{\Lambda}=\frac{\rho_{\Lambda}}{\rho_{c}}=-\nu\approx0.8
\end{equation}
which is not far from the observations. The fraction of the ``dark
energy'' is just the -log of the relative volume at IR limit, which
is always less than 1 since the volume is always shrinking when the
curvature of the universe is positive.

It is worth stressing that in this theory the zero-point quartic divergent
contributing to the cosmological constant does not appear in the effective
action of gravity. The reason is simple, the existence of the zero-point
energy in quantum mechanics, is based on the assumption that there
correspondingly exists a classical, absolute, external and infinitely
precise time parameter known as the Newton's time. And the existence
of the zero-point vacuum energy density in quantum fields theory,
is base on the assumption that there correspondingly exists a classical,
absolute, external and infinitely precise inertial frame $(x,y,z,t)$
known as the Minkowski's spacetime. The parameter background, no matter
the Newton's time or Minkowski's spacetime, is free from any quantum
fluctuations and hence can not be realized precisely. Thus we consider
the \textit{absolute} zero-point vacuum energy corresponding to the
absolute parameter background is unphysical and unobservable including
the Casimir effect \citep{PhysRevD.72.021301}, which can be viewed
as a relative measurement of zero-point fluctuation w.r.t. the external
plates but absolute. This kind of unphysical contribution is not a
severe problem when we are working in an inertial frame (no matter
be Galilean or Lorentz invariant). However, when the general covariance
and gravitational effect are seriously taken into account, it leads
to severe problem. Because the Equivalence Principle asserts that
all kinds of energies must universally gravitate, and we even know
that the electron self-energy coming from the vacuum polarization
(measured by the famous Lamb shift) contributes to its inertial and
gravitate normally \citep{Polchinski:2006gy}. We still have no reason
and evidence that the Equivalence Principle fails at the quantum level.
Here if we replace the Newton's parameter time or Minkowski's inertial
parameter background by the frame fields as a physical reference which
is also subject to the quantum fluctuating, the zero-point energy
density w.r.t. to the physical reference frame fields does not appear
any more, and the leading vacuum energies contribution to the gravitational
effect is the 2nd central moment quantum fluctuation of the frame
fields $\langle\delta X^{2}\rangle$ or $\langle\delta g_{\mu\nu}^{(2)}\rangle$
which not only drives the deformation of the spacetime but also leads
to a correct cosmological constant. Thus in this sense, we consider
the Equivalence Principle can be retained and generalized to the quantum
level.

Another important difference between the effective gravity and the
Einstein' gravity is that the critical density $\lambda=\rho_{c}$,
as the only input coupling, plays the fundamental role of the theory,
while the Newton's constant (no matter in Newton's or Einstein's gravity)
plays no role in it. In the generalized Einstein's gravity with a
cosmological constant, there are two fundamental constants, the Newton's
constant and the cosmological constant, and hence can be seen has
two distinct characteristic scales, the Planck and the Hubble scale.
However, the effective gravity theory here, these two constants combine
into a single constant $\lambda$ and hence the theory allows the
energy scale goes beyond the individual Planck scale ($k\rightarrow\infty$)
and individual Hubble scale ($k\rightarrow0$). Because the characteristic
energy scale related to the critical density $\lambda$ is low, so
the high energy limit of this gravity is trivial while the low energy
long distance behavior is strongly modified, which is not only a possible
solution to the ``dark energy'' puzzle discussed in the next section,
but also possibly related to other long distance anomalies in observations,
e.g. the ``dark matter'' puzzles \citep{Luo:2019iby}.

\section{Observational Effect}

As mentioned in the previous section, the Ricci flow has two effects
on the spacetime: the first one is that the Ricci flow is a coarse
graining process which gradually smoothes out the local inhomogeneous
making the universe more and more homogeneous and isotropic as the
cosmological principle asserts; the second one is that the Ricci flow
continuously deforms the homogeneous and isotropic shrinking soliton
spacetime, from unit volume converge to a finite volume ratio $e^{\nu(M_{\infty})}$
when the Ricci flow is normalized by a counter term playing the role
of a cosmological constant. 

For the spacetime is homogeneous and isotropic so that the space and
time are on an equal footing, the flow effect is an important contribution
to the scale factor and redshift in cosmic observations. Now consider
the squared scale factor $a_{\tau}(T)$ w.r.t. the current epoch one
$a_{\tau}(T_{0})$, both of which are at the same cutoff scale $\tau$,
\begin{equation}
\langle a_{\tau}(T)-a_{\tau}(T_{0})\rangle^{2}\equiv\langle\Delta a_{\tau}(\varDelta T)\rangle^{2}.
\end{equation}
The squared difference of a homogeneous and isotropic scale factor
at cutoff scale $\tau$ is related to the volume flow (\ref{eq:volume constraint-2})
\begin{equation}
\langle\Delta a_{\tau}(\varDelta T)\rangle^{2}=\langle\Delta a_{0}(\varDelta T)\rangle^{2}+\langle\delta a_{\tau}^{2}\rangle=u_{\tau}^{-\frac{2}{D}}\langle\Delta a_{0}(\varDelta T)\rangle^{2}=e^{-\frac{2}{D}\left(\lambda\tilde{N}-\nu\right)}\langle\Delta a_{0}(\varDelta T)\rangle^{2}.\label{eq:a^2}
\end{equation}
Now it has two contributions, the first one $\langle\Delta a_{0}(\varDelta T)\rangle^{2}$
comes from the Hubble expansion which depends on the physical time
interval $\varDelta T$
\begin{equation}
1+\langle z\rangle=\frac{\langle a_{\tau}(T_{0})\rangle}{\langle a_{\tau}(T)\rangle};
\end{equation}
and the second one $\langle\delta a_{\tau}^{2}\rangle$ is due to
the flow effect reflecting the broadening of the redshift, which is
$\tau$ dependent, 
\begin{equation}
1+\frac{1}{2}\langle\delta z^{2}\rangle=\frac{\langle a_{0}^{2}(T)\rangle}{\langle a_{\tau}^{2}(T)\rangle}.
\end{equation}
It is such contribution giving rise to the accelerating expansion.
Note that 
\begin{equation}
\frac{\langle\delta z^{2}\rangle}{\langle z\rangle^{2}}=-2\frac{\langle a_{\tau}^{2}(T)-a_{0}^{2}(T)\rangle}{\langle a_{0}(T)-a_{0}(T_{0})\rangle^{2}}=-2\frac{\langle\delta a_{\tau}^{2}\rangle}{\langle\Delta a_{0}(\varDelta T)\rangle^{2}},
\end{equation}
so at the cosmic scale $\tau\rightarrow0$, $\tilde{N}\rightarrow0$,
from (\ref{eq:a^2}) and (\ref{eq:e^nu}), we have
\begin{equation}
\frac{\langle\delta a_{0}^{2}\rangle}{\langle\Delta a_{0}(\varDelta T)\rangle^{2}}=e^{\frac{2}{D}\nu}-1\approx-0.34
\end{equation}
and hence
\begin{equation}
\frac{\langle\delta z^{2}\rangle}{\langle z\rangle^{2}}\approx0.68.\label{eq:dz2/z2}
\end{equation}

It is worth stressing that at the cosmic scale the variance of the
redshift $\langle\delta z^{2}\rangle$ can not be ignored which is
of order $O(1)$ w.r.t. the squared redshift. The variance of the
redshift gives a correction to the distance-redshift relation at order
$O(z^{2})$ which is significant at large redshift observation. By
expanding the distance by the power of the redshift and using the
relation $\langle z^{2}\rangle=\langle z\rangle^{2}+\langle\delta z^{2}\rangle$,
we have a modified distance-redshift relation
\begin{equation}
\langle d_{L}(z)\rangle=\frac{1}{H_{0}}\left[\langle z\rangle+\frac{1}{2}\langle z^{2}\rangle+O(z^{3})\right]=\frac{1}{H_{0}}\left[\langle z\rangle+\frac{1}{2}\left(1-q_{0}\right)\langle z\rangle^{2}+O(z^{3})\right]
\end{equation}
where $\langle d_{L}(z)\rangle$ is the measured distance between
e.g. supernovas at large redshift and an observer in a lab, and the
deceleration parameter is
\begin{equation}
q_{0}=-\frac{\langle\delta z^{2}\rangle}{\langle z\rangle^{2}}\approx-0.68\label{eq:q0}
\end{equation}
consistent with the observations \citep{Perlmutter:1998np,Riess:1998cb,Ade:2015xua}.

The universality and isotropic of the quantum variance of the redshift
is an indication that the Equivalence Principle could be valid and
generalized to the quantum level. Gravity is not only universally
depicted by the 1st moment of the metric (a universal Hubble's expansion
rate $H_{0}$) but also by the 2nd moment (a universal deceleration
parameter $q_{0}$). The spectral lines taking different energies
are universally free falling, not only be universally redshifted but
also universally be broaden by the quantum variance.

The spacetime coordinates become more and more fuzzy as the Ricci
flow driving the 2nd central moment of the spacetime geometry becomes
more and more important at the cosmic scale. The quantum variance
of the redshift as physical observable becomes more and more unignorable
especially at large redshift regime, making the universe seem accelerating
expansion. In fact, we have not directly measured the 2nd moment fluctuation
of the redshift, instead of measuring the modified distance-redshift
relation at quadratic order $O(z^{2})$. So a direct test of the theory
is to directly measure the intrinsic quantum broadening of the redshift
$\langle\delta z^{2}\rangle$, and check the almost linear dependence
between the quantum broadening and the squared-mean redshift $\langle z\rangle^{2}$,
given by (\ref{eq:q0}), and check whether the proportional constant
is close to the deceleration parameter. Indeed, there are many non-quantum
origins and effects to the dirty variance of the redshift, such as
the thermodynamic broadening. But if one notes that as the distance
scale becomes larger and larger, the ratio (\ref{eq:q0}) becomes
of $O(1)$, that is to say, the quantum part of the variance may become
more and more dominant. On the other hand, unlike other non-quantum
effects, the quantum part of the variance is universal and isotropic
as the Equivalence Principle claims which may differ it from other
non-quantum noises. Thus it may be feasible to measure the intrinsic
quantum variance and its ratio to the squared redshift.

From this point of view, the ``cosmic coincidence problem'' or ``why
now problem'' can be put back into perspective. The problem wonders
why the ``dark energy'' density is comparable to the matter density
or critical density ``now'', if the ``dark energy'' density and
the matter density are redshifted in complete different ways and hence
should only be comparable in a special epoch. Here it is reasonable
to see that we are in fact not in a special moment in the history
of the universe, since no matter when one performs a cosmic observation,
the ``dark energy'' as a ``mirage'' coming from the broadening
of the redshift (distant clock) w.r.t. the one in the observer's epoch
(fiducial lab frame) is ``always'' seen being of order of the critical
density, $\Omega_{\Lambda}\sim O(1)$. Thus if the matter density
fraction is relatively redshifted as $\Omega_{m}(1+z)^{3}$ from ``now''
(i.e. w.r.t to the fiducial lab frame which could be at any moment
in the history of the universe), the ``dark energy'' is ``always''
comparable with the matter density at the ``coincident'' redshift
$z_{c}=\left(\frac{\Omega_{\Lambda}}{\Omega_{m}}\right)^{1/3}-1\sim O(1)$. 

\section{Conclusions}

In this paper, we propose and review a quantum fields theory of spacetime
reference frame, and consider it is a possible theory of quantum spacetime.
The 2nd order central moment quantum fluctuation of the frame fields
induces the Ricci flow of the spacetime. The theory is quantized by
the functional integral method, and via the change of the functional
integral measure we deduce the trace anomaly of the theory which is
because the Ricci flow in nature does not preserve the volume of the
spacetime even if the volume of the lab could be fixed. We argue that
the effect of the volume change of spacetime (due to 2nd moment fluctuation
and Ricci flow) can be directly tested by observing the variance of
the redshift which plays the role of a ruler or clock at distance.
We find that the trace anomaly is closely related to some of Perelman's
functionals: a formal trace anomaly is given by the Shannon entropy
of the spacetime with an unnormalized density, a normalized trace
anomaly which is vanished at the final IR limit is just the -log of
the Perelman's partition function. The cancellation of the normalized
trace anomaly in a lab up to high energy UV scale is considered as
a consistency condition for a general covariant theory at the quantum
level. The counter term at UV is given by the UV limit of the W-functional
or log of the reduced volume of a maximally symmetric shrinking Ricci
soliton, which correctly leads to the emergence of the cosmological
constant. The theory makes the quartic zero-point divergent contribution
to the cosmological constant dissapear, and the leading vacuum energies
contribution to the cosmological constant and gravitational effect
is the 2nd central moment fluctuation of the frame fields, which gives
a value consistent with observation: it is of order of the critical
density $\lambda$ (as the only input of the theory) and has a fraction
$\Omega_{\Lambda}\sim0.8$. For this reason, we assume the validity
of the Equivalence Principle still holds at the quantum level. A modified
distance-redshift relation is calculated within the framework giving
a deceleration parameter $q_{0}\approx-0.68$. 
\begin{acknowledgments}
This work was supported in part by the National Science Foundation
of China (NSFC) under Grant No.11205149, and the Scientific Research
Foundation of Jiangsu University for Young Scholars under Grant No.15JDG153.
\bibliographystyle{plain}

\end{acknowledgments}

\end{document}